\input harvmac.tex
%
\figno=0
\def\fig#1#2#3{
\par\begingroup\parindent=0pt\leftskip=1cm\rightskip=1cm\parindent=0pt
\baselineskip=11pt
\global\advance\figno by 1
\midinsert
\epsfxsize=#3
\centerline{\epsfbox{#2}}
\vskip 12pt
{\bf Fig. \the\figno:} #1\par
\endinsert\endgroup\par
}
\def\figlabel#1{\xdef#1{\the\figno}}
\def\encadremath#1{\vbox{\hrule\hbox{\vrule\kern8pt\vbox{\kern8pt
\hbox{$\displaystyle #1$}\kern8pt}
\kern8pt\vrule}\hrule}}

\overfullrule=0pt

%

\def\np#1#2#3{{\it Nucl. Phys.} {\bf B#1} (#2) #3}
\def\pl#1#2#3{{\it Phys. Lett. }{\bf B#1} (#2) #3}
\def\prl#1#2#3{{\it Phys. Rev. Lett.}{\bf #1} (#2) #3}
\def\physrev#1#2#3{{\it Phys. Rev.} {\bf D#1} (#2) #3}

\font\zfont = cmss10 

\def\bigone{\hbox{1\kern -.23em {\rm l}}}
\def\ZZ{\hbox{\zfont Z\kern-.4emZ}}

\def\a{\alpha}
\def\b{\beta}
\def\g{\gamma}
\def\d{\delta}

\def\m{\mu}
\def\n{\nu}

\def\s{\sigma}

\def\ps{\psi}
\def\G{\Gamma}
\def\D{\Delta}

\def\o{\over}

\def\partialsl{\partial \!\!\! /}

\Title{
{\vbox{
\rightline{\hbox{hepth/0010282}}
\rightline{\hbox{CALT-68-2302}}
\rightline{\hbox{UMD-PP-01-019}}
}}}
{\vbox{\hbox{\centerline{Compactifying 
${\cal M}$-Theory to Four Dimensions}}
}}
\smallskip
\centerline{Katrin Becker\footnote{$^1$}
{\tt beckerk@theory.caltech.edu}} 
\smallskip
\centerline{\it California Institute of Technology 452-48, 
Pasadena, CA 91125}
\centerline{\it CIT-USC Center for Theoretical Physics}
\smallskip
\centerline{Melanie Becker\footnote{$^2$}
{\tt melanieb@physics.umd.edu} }
\centerline{\it Department of Physics, University of Maryland}
\centerline{\it College Park, MD 20742-4111}

\bigskip
\bigskip
\baselineskip 18pt
\bigskip
\noindent

We consider compactifications of ${\cal M}$-theory to four-dimensional 
Minkowski space on seven-dimensional
non-compact manifolds.
These compactifications include a warp factor which is non-constant 
due to the presence of sources coming from
fivebranes wrapping two-dimensional submanifolds of the internal
seven-dimensional space. We derive the expression
for the field strengths and consider an explicit example
of this general class of solutions.

\Date{October, 2000}

\newsec{Introduction}

Recently warped compactifications of ${\cal M}$-theory and ${\cal F}$-theory 
have become a 
fascinating arena of research because among other things it was suggested in 
\ref\cpv{C.~S.~Chan, P.~L.~Paul and H.~Verlinde,``A Note on 
Warped String Compactification'', \np {581} {2000}  {156}, hep-th/0003236.}
that 
these compactifications may
provide us with a string theoretic realization of 
the Randall-Sundrum (RS) scenario
\ref\rs{L.~Randall and R.~Sundrum, ``A Large Mass 
Hierarchy from a Small Extra
Dimension'', \prl {83} {1999} {3370}, hep-th/9905221.}, 
\ref\rs{L.~Randall and R.~Sundrum, ``An Alternative to 
Compactification'', 
\prl {83} {1999} {4690}, hep-th/9906064.}. 
Using a metric with a warp factor Randall and Sundrum argued 
that our four-dimensional world could be described as the world-volume 
of a `threebrane'. 
In the five-dimensional example suggested by Randall and Sundrum gravity is 
localized on the four-dimensional `brane' even if the fifth dimension 
is infinitely extended. This was only possible because of 
an exponential warp factor in the metric.

Warped compactifications have been known in string theory for a long time
(see e.g.
\ref\wnw{B.~De Wit, H.~Nicolai and N.~P.~Warner, ``The embedding of Gauged
$N=8$ Supergravity into $d=11$ Supergravity, \np {255} {29} {1985}.}, 
\ref\str{A.~Strominger, ``Superstrings with Torsion'', \np {274} {253} {1986}.}
and 
\ref\wsd{B. de Wit, D.~J.~Smit and N.~D.~Hari Dass, ``Residual Supersymmetry
of Compactified $D=10$ Supergravity'', \np {283} {1987} {165}.}) and thus string
theory 
seems to be the natural framework in which the RS-scenario
could be realized.
Indeed, it was suggested in {\cpv} that compactifications of ${\cal F}$-theory 
on an elliptically
fibered Calabi-Yau four-fold 
\ref\svw{S.~Sethi, C.~Vafa and E.~Witten, ``Contraints on Low Dimensional
String compactification'', \np {480} {1996} {213}, hep-th/9606122.}
provide a realization of the 
RS-scenario and the consistency conditions following from
supersymmetry can be obtained from the ${\cal M}$-theory compactifications
on four-folds considered in 
\ref\bb{K.~Becker and M.~Becker, ``${\cal M}$-Theory on Eight Manifolds'', 
\np {477} {1996} {155}, 
hep-th/9605053.}.

Warped compactifications play also an important role in the description 
of confining supersymmetric gauge theories and ultimately 
in the description of QCD. 
This is because there is a close relation between warped 
compactifications and 
Ramond-Ramond backgrounds in string theory.
Confining 
gauge theories can be realized, for example, as perturbations 
by 3-form flux of type IIB string theory
on $AdS_5 \times S_5$. In this case the resulting supergravity theory
has a naked singularity and it was shown by Polchinski and Strassler 
\ref\ps{J.~Polchinski and M.~J.~Strassler, ``The String Dual of a Confining
Four-Dimensional Gauge theory'', hep-th/0003136. } 
that this singularity actually corresponds to an expanded
brane source. The 3-form flux of the supergravity theory corresponds
to a perturbation of the ${\cal N}=4$ gauge theory by mass terms and the
resulting gauge theory has ${\cal N}=1$ supersymmetry {\ps}. 

Non-perturbatively, ${\cal N}=1$ supersymmetric gauge theories 
can be realized by placing D3-branes at 
conical singularities of a Ricci-flat six-dimensional cone
whose base manifold is a five-dimensional Einstein space $X_5$. 
On the supergravity side one considers type IIB theory 
on $AdS_5 \times X_5$ and this is dual to the world-volume 
theory of the D3-branes at the singularity. In case that 
one considers D3-branes on the conifold
\ref\kw{I.~R.~Klebanov and E.~Witten, ``Superconformal Field Theory on
Threebranes at a Calabi-Yau 
Singularity'', \np {536} {1998} {199}, hep-th/9807080.}
, for example, one would obtain 
on the worldvolume of the D3-branes a gauge theory with $SU(N) \times SU(N)$  
gauge group. 
Besides considering D3-branes it is also possible to consider 
D5-branes wrapped on collapsed 2-cycles at the singularity
\ref\kt{I.~R.~Klebanov and A.~A.~Tseytlin, ``Gravity Duals of Supersymmetric
$SU(N) \times SU(N+M)$ Gauge Theories'', \np {578} {2000} {123}.}.
This has the effect that the D3-brane charge eventually becomes negative and the
supergravity metric becomes singular. 
It was argued by Klebanov and Strassler
\ref\ks{I.~R.~Klebanov and M.~J.~Strassler, ``Supergravity and a Confining
Gauge Theory: Duality Cascades and $\chi$SB-Resolution of Naked Singularities'',
JHEP {\bf 0008} (052) 2000, hep-th/0007191.} that
this naked singularity of
the metric gets resolved in terms of a warped deformed conifold 
which is completely non-singular. 
It was realized recently in
\ref\gp{M.~Gra\~na and J.~Polchinski, ``Supersymmetric 3-Form Flux and
Perturbations of $AdS(5)$'', hep-th/0009211.} and 
\ref\gub{S.~S.~Gubser, ``Supersymmetry and F-Theory Realization 
of the Deformed Conifold with 3-Form Flux'', hep-th/0010010. } 
that the Klebanov-Strassler model can be obtained as a special  
case of the solutions derived in {\bb} describing compactifications
of ${\cal M}$-theory on eight-manifolds. This is interesting and 
one may wonder if there is a similar connection between the
models considered in {\bb} or a corresponding generalization thereof
and the Polchinski-Strassler model. This would be useful
to derive the exact solution 
of the model considered in {\ps}.

In this paper we would like to broaden the class of theories 
that admit non-vanishing tensor fields. 
We would like to consider compactifications of 
${\cal M}$-theory to four-dimensional Minkowski space on 
seven-dimensional non-compact manifolds.  
These compactifications 
will involve a warp factor and we will see that we are able to 
construct solutions with non-vanishing expectation
values for the tensor fields and a non-constant warp factor 
while supersymmetry is being preserved.

It has been known for some time that supersymmetry requires that 
compactifications of
eleven-dimensional supergravity to four-dimensional
Minkowski space have vanishing expectation values for the 
4-form field strength and a constant warp factor if the internal  
manifold is compact and no sources are being considered
\ref\cara{P.~Candelas and D.~J.~Raine, ``Spontaneous Compactification 
and Supersymmetry in $D=11$ Supergravity'', \np {248} {1984} {248}.}, 
{\wsd}.
 
But the argument presented in the previous two papers 
fails if the internal manifold is non-compact 
or sources are considered as we will see in this paper. 
We shall see that in this case one is able to preserve supersymmetry
and obtain non-vanishing expectation values for tensor fields and a 
non-constant warp factor at the same time.
The sources that we shall be interested in originate from ${\cal M}$-theory
fivebranes wrapping two-dimensional submanifolds of the internal 
seven-dimensional space.
This has the consequence that the internal space is non-compact.
We will derive the explicit form for the field strength that follows from 
supersymmetry.

An interesting example of the general class of solutions 
that we will find here
was
considered in
\ref\fs{A.~Fayyazuddin and  D.~J.~Smith, ``Localized Intersections
of $M5$-Branes and Four-Dimensional Superconformal Field Theories'', 
JHEP {\bf 9904} (030) 1999, hep-th/9902210.} and
\ref\bfms{B.~Brinne, A.~Fayyazuddin, S.~Mukhopadhyay and D.~J.~Smith, 
``Supergravity $M5$-Branes Wrapped on Riemann Surfaces and Their
QFT Duals'', hep-th/0009047.} where the seven-manifold is a
warped product of a four-dimensional K\"ahler manifold times a 
three torus. In this example the 
${\cal M}$-theory fivebranes are wrapped on 2-cycles of the four-dimensional
K\"ahler manifold.
We will show explicitly that this example solves the equation for the
field strength obtained herein.

This paper is organized as follows. 
In section 2 we consider compactifications of 
${\cal M}$-theory to four-dimensional
Minkowski space on non-compact seven-manifolds. 
In section 2.1 we review the argument
which shows why the field strengths are vanishing and 
the warp factor is constant
for compactifications of ${\cal M}$-theory on compact 
seven-manifolds.
In section 2.2 we consider non-compact seven manifolds 
where sources coming
from fivebranes wrapping two-dimensional submanifolds 
of the internal space are taken into account.
We derive the expression for the field strength in four dimensions 
that follows from supersymmetry and show that the warp
factor is non-constant due to the presence of these sources. 
In section 3 we show that the compactifications considered by 
Fayyazuddin and Smith in {\fs} and {\bfms} provide an explicit solution 
of our equations.
Some concluding remarks are made in section 5 and in the appendix we list
some relevant formulas.

\newsec{${\cal M}$-theory Compactifications to d=4 Minkowski Space}
In this section we would like to consider compactifications of 
eleven-dimensional supergravity on non-compact seven-manifolds
and derive the explicit form of the 4-form field strength that 
follows from supersymmetry\foot{For recent work on compactifications on 
compact seven-dimensional 
manifolds see 
\ref\gukov{S.~Gukov, ``Solitons, Superpotentials and Calibrations'', 
\np {574} {169} {2000}, hep-th/9911011.}, 
\ref\acha{B.~S.~Acharya and B.~Spence, ``Flux, Supersymmetry and M-Theory 
on Seven Manifolds'', hep-th/0007213. } and 
\ref\hern{R.~Hernandez, ``Calibrated Geometries and Nonperturbative 
Superpotentials in M-theory'', hep-th/9912022.}.}.
\subsec{Compact Seven-Manifolds}
Let us first start by reviewing the argument that leads to the conclusion 
that supersymmetry implies that the warp factor is constant and 
thus the field strengths have to 
vanish for compact seven-manifolds {\cara}, {\wsd}.
The bosonic part for eleven-dimensional supergravity Lagrangian 
contains a 3-form $C$ with field strength $F$ and the dual 7-form
$\star F$
\eqn\aaai{
{\cal L}= { 1\over \kappa^2} \int d^{11} x \sqrt{g} \left( 
-{1 \over 2} R -{1 \over 48} F_{IJKL} F^{IJKL} 
-{\sqrt{2} \over 3456} \epsilon^{I_1 I_2 \dots I_{11}}
C_{I_1 I_2 I_3} F_{I_4 \dots I_7} F_{I_8 \dots I_{11}}\right).
}
The complete action
is invariant under the supersymmetry transformations
\eqn\aviii{
\eqalign{
\delta {e_I}^m & ={ 1\over 2} {\bar \eta} \Gamma^m \Psi_I , \cr 
\delta C_{IJK} & =- {\sqrt{2} \over 8} {\bar \eta} \Gamma_{[ IJ}
\Psi_{k]},  \cr 
\d \Psi_M & = \nabla_M\eta +{\sqrt{2} \over 288} 
({\G_M}^{PQRS}-8 \d_M^P \Gamma^{QRS} )F_{PQRS} \eta.\cr}
}
The Einstein equation following from {\aaai} takes the form 
\eqn\aii{
R_{MN}-{1 \over 2} g_{MN} R+T_{MN}=0,
}
where $T_{MN}$ is the energy-momentum tensor of the 4-form field strength 
$F$ given by
\eqn\aiii{
T_{MN}=4F_{MPQR} {F_{N}}^{PQR}-{1 \over 2} g_{MN} F_{PQRS}F^{PQRS}.
}
In eleven dimensions one could in principle have membranes and fivebranes 
that couple to the action {\aaai} and appear as sources in the equation
of motion for $C$ and Bianchi identify respectively. 
Ignoring (for the moment) the presence
of these sources $C$ satisfies the equation of motion 
\eqn\aiiixxi{
d\star F=-{1 \over 2} F \wedge F, 
}
and the Bianchi identity
\eqn\aiiixx{
dF=0.
}
We now would like to consider a line element of the form
\eqn\ai{
ds^2=g_{MN} dx^M dx^N= 
\Delta^{-1}(y) \eta_{\mu \nu} dx^{\mu} dx^{\nu} 
+g_{mn}(y) dy^m dy^n. 
}
Here $\Delta(y)$ is the warp factor that because of Poincar\' e invariance 
depends only on the coordinates of the internal manifold,
$g_{mn}$ is the metric of the internal seven-dimensional space and
$\eta_{\mu \nu}$ is the four-dimensional Minkowski space metric.

In compactifications with a maximally symmetric four-dimensional space-time
the non-vanishing components of the 4-form field strength are
\eqn\aixi{
\eqalign{
& F_{mnpq}, \cr
& F_{\mu \nu \rho \sigma} =f \epsilon_{\mu \nu \rho \sigma}.\cr }
}
Here $f$ is arbitrary and will be determined later on and 
$\epsilon_{\mu \nu \rho \sigma}$ is the antisymmetric tensor of four-dimensional
Minkowski space.
Taking into account the decomposition {\ai} of the metric and {\aixi} one obtains two
equations from the eleven-dimensional Einstein equation{\aii}, 
one for the external component (i.e. where the Ricci-tensor
has four-dimensional Minkowski indices) and one for the internal
component (where the Ricci-tensor has seven-dimensional indices).
The equation for the external component leads to the following equation
\eqn\aiv{
\nabla^m (\Delta^{-3} \partial_m \Delta ) 
=-{ 2 \over 3} \Delta^{-2} (F^2 +48 {\Delta}^4 f^2),
}
where the covariant derivative involves the Christoffel connection as usual.
Using Stokes theorem we see that the integral of the left-hand side 
of this equation over a compact manifold vanishes. 
To conclude that all the expectation values of $F$ actually vanish 
one uses the observation that the right hand side of equation {\aiv}
is negative. Therefore all the components of $F$ must vanish
\eqn\av{
F_{PQRS}=0.
}
Going back to equation {\aiv} one obtains the Laplace equation
\eqn\avi{
\nabla^m(\Delta^{-3} \partial_m \Delta )=0,
}
whose only solution on a compact manifold is
\eqn\avii{
\D=const.
}
Therefore, we recover the conventional supergravity compactifications in which no 
warp factor was taken into account. 

\subsec{Non-compact Seven-Manifolds}

However, the argument of the previous section fails in a rather interesting way 
for non-compact internal seven-manifolds where, for example,  
${\cal M}$-theory fivebranes that wrap cycles of the seven-manifold are
taken into account.
For non-compact manifolds the integral over the left hand side of
{\aiv} is not equal to zero because of boundary terms. 
Furthermore, external sources 
modify the right hand side of this equation because these sources contribute to
the energy momentum tensor {\aiii} and therefore to the Einstein equation {\aii}.  
Very generally, we will consider seven-manifolds having a 
two-dimensional submanifold 
on which the fivebrane can be wrapped. The fivebrane worldvolume   
is of the form $M_4 \times \Sigma$, where $M_4$ is four-dimensional
Minkowski space and $\Sigma$ the two-dimensional submanifold of the 
seven-dimensional internal space. 
As we will see in the 
following, in this situation it is possible to find non-vanishing 
expectation values for antisymmetric tensor fields with 
unbroken supersymmetry and a non-constant warp factor.

To find these solutions we will perform a similar analysis as in {\bb} 
but now for non-compact seven-manifolds. 
Unbroken supersymmetry requires that the transformations
{\aviii} are zero. The first two equations are satisfied because
in the classical background the gravitino vanishes.   
We therefore only have to consider the supersymmetry 
transformation\foot{We will be following 
the notation and conventions of {\wsd}.} of the eleven-dimensional
gravitino {\aviii}.
The condition for unbroken supersymmetry 
\eqn\aaxivii{\delta \Psi_M=0,} 
will be decomposed into the external and internal components as follows
\eqn\aix{
\eqalign{
& \left(\nabla_{\mu} + \D^{-1/2} \g_\mu T
\right) \eta=0, \cr
& \left(\nabla_m +O_m\right) \eta =0.\cr }
}
Here we have defined the quantities
\eqn\ax{
\eqalign{
&T= {\sqrt{2} \o 288}(F -i f \D^2 \gamma_5 -
36 \sqrt{2} \g_5  \partialsl \log \Delta ),\cr 
&O_m={\sqrt{2} \o 288}({ 1\o 2 } i f \gamma_m +\g_5 (\g_m F -12 F_m )). \cr}
}
Furthermore, we have used the notation 
\eqn\axii{
\eqalign{
& F=\gamma^{pqrs} F_{pqrs}, \cr 
& F_m =\gamma^{pqr} F_{mpqr},\cr}
}
and $\partialsl=\g^m \partial _m$.
The eleven-dimensional gamma matrices have been 
decomposed into two sets of mutually commuting gamma matrices according to
\eqn\aaxv{
\eqalign{
\Gamma_{\mu} & =\gamma_{\mu} \otimes 1 , \cr
\Gamma_m & =\gamma_5 \otimes \gamma_m ,\cr}
}
which is appropriate for an $11=4+7$ split.
Furthermore, we have chosen our gamma matrices to be hermitian 
and ${\gamma_5}={\gamma_1} {\gamma_2} {\gamma_3}{\gamma_4}$
satisfies $(\gamma_5)^2=1$.
We will decompose the eleven-dimensional spinor $\eta$ according
to
\eqn\aaxiv{\eta=\epsilon \otimes \xi,}
where $\epsilon$ is a four-dimensional anticommuting spinor, while $\xi$ 
is a commuting seven-dimensional Majorana spinor. Without loss of generality
we will consider ${\epsilon}$ to be a positive chirality spinor 
${{\gamma}_5}{\epsilon}={\epsilon}$.  

Since we are considering compactifications to four-dimensional Minkowski space
we set 
\eqn\aaxivix{\nabla_\mu \epsilon=0,}
which for maximally symmetric four-dimensional space-time implies 
that the external space-time is flat Minkowski. 
Therefore, from the external component of the gravitino transformation
(i.e. the first equation in {\aix}) one obtains the expression
\eqn\axiii{(F -i f \D^2 \gamma_5 -
36 \sqrt{2} \g_5  \partialsl \log \Delta){\eta} =0.
}
Since the gamma matrices are hermitian one concludes from {\axiii}
\eqn\axiv{
f=0.
}
Therefore, unbroken 
supersymmetry does not allow external components $F_{\m\n\rho\s}$ 
for compactifications to
four-dimensional Minkowski space independently if the internal manifold is 
compact or not. 
However, the situation is different for the internal 
components $F_{mnpq}$ which are now constrained to satisfy 
\eqn\axv{
F \xi = 36 \sqrt{2} 
\partialsl \log \Delta \xi.
}
Notice that for compactifications of ${\cal M}$-theory 
on seven-dimensional manifolds the
equations satisfied by the field strengths are rather different 
than for compactifications of ${\cal M}$-theory
on eight manifolds considered in {\bb}. In {\bb} it was found that
the internal components of $F$ satisfy the equation
\eqn\axvii{
F \xi = 0,
}
rather than {\axv}. Furthermore it was found in {\bb} that the external component of
$F$ could be expressed 
in terms of a derivative of the warp factor rather than having to vanish as
in {\axiv}.

Using the equations {\axiv} and {\axv} for the field strength
it is possible to rewrite 
the supersymmetry transformations of the gravitino {\aix} in the following form
\eqn\axvii{
\eqalign{ & {\tilde \nabla}_m{\tilde \xi} -{\sqrt{2}\over 24} \Delta^{3/2} F_{mpqr} 
{\tilde \gamma}^{pqr} {\tilde \xi}  =0,\cr 
& F_{pqrs} {\tilde \gamma}^{pqrs} {\tilde \xi} =36\sqrt{2} \Delta^{-5/2} 
\partial_a \Delta {\tilde \gamma}^a {\tilde \xi}.\cr}
}
Here we have introduced a new metric ${\tilde g}_{mn}$ which is related 
to the metric $g_{mn}$ appearing in {\ai} by a rescaling with the warp factor
\eqn\axviii{
{\tilde g}_{mn}=\Delta g_{mn}.
}
The gamma matrices are rescaled accordingly ${\tilde \gamma}_m=
\Delta^{1/2} \gamma_m$. 
We have also rescaled the seven-dimensional spinor $\xi$ 
\eqn\axix{
{\tilde \xi} =\Delta^{1/4} \xi, 
}
and used an identity relating covariant derivatives of spinors
with respect to conformally transformed metrics that we have included
in the appendix.
The first relation in {\axvii} guarantees that we can find a spinor 
whose norm is covariantly constant
\eqn\axx{
{\tilde \nabla }_m ( {\tilde \xi}^{\dagger} {\tilde \xi})  =0, 
}
and therefore we can choose the normalization 
\eqn\axxi{
{\tilde \xi}^{\dagger} {\tilde \xi} =1. 
}
In terms of this spinor we can define a 2-form as
\eqn\axxii{
{\tilde \omega}_{ab} = i {\Delta}^{-3} {\tilde \xi}^{\dagger} 
{\tilde \gamma}_{ab} {\tilde \xi},
}
where we have introduced the warp factor for convenience.
We shall see in a moment that the tensor field can be expressed 
in terms of this 2-form.
In general, seven-dimensional manifolds are not characterized by 2-forms. 
So for example, a $G_2$ holonomy manifold is characterized 
by a 3-form $\Phi$ and 
it's Hodge dual 4-form $\star \Phi$ and not by a 2-form. 
But we can still expect to find seven-dimensional
manifolds
with non-vanishing 2-forms in special cases. Precisely these manifolds 
will be the interesting ones for which the field strengths have non-vanishing
expectation values and the warp factor is non-trivial.
In the next section we will consider
a seven-manifold that is a warped product of a four-dimensional K\"ahler manifold
times a 3-torus. In this case the above 2-form is related to the 
K\"ahler form of the four-dimensional K\"ahler manifold. 
This type of compactification
was considered in {\fs} and {\bfms}. 
In this section we will derive the explicit expression for the 
expectation value of the tensor field.
A more detailed analysis of the properties of the
background geometry will appear elsewhere
\ref\toa{Work in progress.}.

In eleven dimensions a 4-form $F$ is dual to a 7-form $\star F$.
When compactifying on a seven-manifold to four-dimensional 
Minkowski space we can define a 3-form field strength $K$ 
in the following way
\eqn\bxxi{
\star F = V_4 \wedge K . 
}
Here $ V_4=dx^0 \wedge dx^1 \wedge dx^2 \wedge dx^3$ 
is the constant volume element of the four-dimensional 
Minkowski space and $\star$ is the Hodge dual with respect to the 
eleven-dimensional metric $g_{MN}$. 
Using some gamma matrix identities that we list in the appendix 
and formulas {\axvii} and {\axxii} it is 
possible to show that $K$ can be expressed through the 
derivatives of $\tilde \omega_{ab}$ 
in the following way: 
\eqn\axxiix{
K_{abc} ={ 3 \over \sqrt{2}} {\tilde \nabla} _{\lbrack a} 
{\tilde \omega}_{bc\rbrack}.
}
To derive this equation it is useful 
to take the identity
\eqn\aaxxiii{
{\tilde \xi}^{\dagger}\{ T, {\tilde \gamma}_{abc} \} {\tilde \xi} =0, 
}
into account. Here $T$ is defined as in {\ax} with $f=0$.
Remembering that the Christoffel connection $\Gamma_{MN}^P$ is 
symmetric in it's lower indices we see that $K$ can be written in the form
\eqn\aaxxii{
K={1 \over \sqrt{2}} d {\tilde \omega} .
}

This is the general solution for $K$ that follows from supersymmetry.
It is rather interesting that the $K$-field can be determined 
explicitly in terms of the 2-form {\axxii} rather than through a 
determining equation as for ${\cal M}$-theory compactifications on 
eight-manifolds (see eqn. (2.52) of {\bb}).
A similar situation appeared in {\str} and 
\ref\vacconf{P. Candelas, G. T. Horowitz, A. Strominger and 
E. Witten, ``Vacuum Configurations for Superstrings'', 
\np {258} {1985} {46}}, where compactifications of the heterotic
string were considered. In {\str} was found that the 
$H$-field could be expressed in terms of a 2-form 
(see eqn. (2.17) of that paper), 
while the Yang-Mills field strength satisfies the 
Donaldson-Uhlenbeck-Yau equation,
which is a determing equation similar as in {\bb}. 
The above result for the $K$-field is reminiscent of the 
generalized calibrations 
considered in \ref\gupa{J. Gutowski and G. Papadopoulos, 
``AdS Calibrations'', \pl {462} {81} {1999}, hep-th/9902034. }, 
\ref\gupato{J. Gutowski, G. Papadopoulos and P. K. Townsend, 
``Supersymmetry and Generalized Calibrations'', \physrev {60} {106006} {1999}, 
hep-th/9905156.} and  
\ref\cekt{H.~Cho, M.~Emam, D.~Kastor and J.~Traschen, ``Calibrations and
Fayyazuddin-Smith Spacetimes'', hep-th/0009062.}.

With the expression  {\aaxxii} for the $K$-field and 
{\bxxi} it is easy to see that the above solution 
satisfies the eleven-dimensional fivebrane Bianchi identity
(or membrane equation of motion) 
\eqn\aaxxiii{
d \star F =0. 
}
Here we have to take into account that because the external components 
of $F$ vanish we have the condition $ F \wedge F=0$. 
Since we are considering fivebrane sources in eleven dimensions,
the fivebrane equation of motion (or membrane Bianchi identity)
is no longer described by the
equation {\aiiixx} but rather by an equation of the form
\eqn\aaxxiv{
dF=\delta.
}
Naivly, the fivebrane sources involve 
delta functions that are supported on the fivebrane worldvolume.
These naive definition, however, leads to inconsistencies of the theory in 
the form of gravitational anomalies and a more careful 
analysis of the right hand side
of equation {\aaxxiv} is in order.
This has been done in
\ref\witten{E.~Witten, ``Five-Brane Effective Action in M-Theory'', 
J.Geo.Phys.{\bf 22} (1997) 103, hep-th/9610234.},
\ref\fhmm{D.~Freed, J.~A.~Harvey, R.~Minasian and G.~Moore, 
``Gravitational Anomaly Cancellation for M Theory Fivebranes'',
Adv. Theor. Math. Phys {\bf 2} (1998) 601, hep-th/9803205.}
and
\ref\bbb{K.~Becker and M.~Becker, ``Fivebrane Gravitational 
Anomalies'',  \np {577} {2000} {156}, hep-th/9911138.}. We will
not enter into a more detailed discussion of these issues here.

Finally, the fivebrane sources will modify the right hand side of Einstein's
equation {\aiv} as the energy momentum tensor of the fivebrane
source has to be taken into account.
Roughly, the external component of this equation takes the form
\eqn\aivxx{
\nabla^m (\Delta^{-3} \partial_m \Delta ) 
=-{ 2 \over 3} \Delta^{-2} F^2 +\delta,
}
where the delta functions come from the fivebrane energy momentum tensor.    
Therefore, if sources are taken into account the warp factor no longer 
satisfies a Laplace equation 
on a compact manifold and the conclusions of section (2.1) are not valid.

At this point we have determined the expectation value for the tensor
field that follows from supersymmetry. Equation {\aivxx} 
is an equation for the
warp factor, while {\aaxxiv} is a determining equation for the 2-form 
{\axxii}. The non-vanishing component of the tensor field then follows
from {\axxiix}. 
The internal component of Einstein's equation will then provide the information
on what the possible seven-dimensional backgrounds are.
A detailed analysis of the possible background geometries will appear 
elsewhere {\toa}. In the next section we shall consider a particular background
geometry.

\newsec{The Fayazzudin-Smith Manifold}
In this section we would like to consider a special seven-dimensional
manifold which is a warped product of a four-dimensional 
K\"ahler manifold times a three-torus. 
We will consider fivebrane sources and the fivebrane wraps 
a 2-cycle of the four-dimensional K\"ahler space. 
This example of seven-manifold gives a supersymmetric 
compactification of ${\cal M}$-theory and was considered 
by Fayyazuddin and Smith in {\fs} and {\bfms}. 
In this case the 2-form ${\tilde {\omega}}$ appearing in {\axxii} and
{\axxiix} 
arises naturally in terms of the K\"ahler form of the 
four-dimensional K\"ahler space. 

The rescaled line element considered in this compactification
is
\eqn\axxiii{
d{\tilde s}^2= \eta_{\mu \nu} dx^{\mu} dx^{\nu} +
2g_{m {\bar n}} dy^m dy ^{\bar n} +
\D^3(y) \d_{\a\b} dy^{\a} dy^{\b}
,}
where $g_{m {\bar n}}$ is the metric of the four-dimensional K\" ahler space 
and $\d_{\a\b}$ is the metric of the three-torus. $\Delta(y)$ is allowed
to depend on both $y^m$ and $y^{\alpha}$.

The 2-form has the several components. The relevant one is 
the K\" ahler form of the four-dimensional manifold ${\tilde \omega} = 
\D^{-3}J$.
Using {\axxii} we can compute $K$ and get
\eqn\axxv{
K= {1 \over \sqrt{2}} d\left( \D^{-3} J \right) . 
}
This is precisely the result obtained in the papers by
Fayyazuddin and Smith (see e.g equations (4)-(6) of {\bfms})
after dualizing and taking a rescaling of 
$F$ into account.
Therefore the tensor fields of the examples considered in {\fs} 
and {\bfms} in which the seven-manifold is a warped product 
of a four-dimensional K\" ahler manifold times a three-torus 
provide an explicit example of the general class 
of solutions described in the previous section.

\newsec{Conclusions and Outlook}

In this paper we have considered warped compactifications of ${\cal M}$-theory
on non-compact seven-manifolds to four-dimensional Minkowski space. 
Conventional compactifications of ${\cal M}$-theory on compact seven-manifolds
lead to vanishing expectation values for the tensor field and a constant
warp factor if supersymmetry 
is imposed. However, this is not the case for the compactifications considered
herein, where it is possible to preserve supersymmetry while the
expectation value of the tensor fields are non-vanishing and the warp factor
is not constant.
This is due to the presence of fivebrane sources which wrap 2-cycles
of the internal seven-dimensional manifold.
We have computed the expression for the field strength 
following from supersymmetry
explicitly. 
Furthermore, we have shown that the compactifications considered in 
{\fs} and {\bfms} provide a concrete example of our general solutions.

It would certainly be interesting to see if further examples that 
solve {\aaxxii} could be found. 
In this paper we have seen how to obtain the form of tensor field 
once the background geometry is known. 
But supersymmetry can, of course, teach us more. A detailed analysis of
the properties of the background geometry will appear elsewhere
{\toa}.

\vskip 1cm

\noindent {\bf Acknowledgement}

\noindent 

Its a pleasure to  H. Verlinde and especially E. Witten 
for discussions and for bringing reference {\bfms} to our attention. 
We also would like to thank A. Strominger for discussions on 
warped compactifications some time ago. 
This work was supported by the U.S. Department of Energy 
under grant DE-FG03-92-ER40701.  
 
\vskip 1cm

{\noindent} {\bf Appendix}

\noindent In this appendix we would like to collect a few useful formulas 
and we would like to explain our notation. 

\item{$\triangleright$} The different types of indices that we use are

\item{} $M,N,\dots$ are eleven-dimensional indices
\item{} $m,n,\dots$ denote seven-dimensional indices 
\item{} $\mu,\nu,\dots$ are the indices of the external space

\item{$\triangleright$} $n$-forms are defined with a factor $1/n!$. 
For example
$$
F={ 1\over 4!} F_{mnpq} dx^m \wedge dx^n \wedge dx^p \wedge dx^q.
$$
\item{$\triangleright$} The gamma matrices $\Gamma_M$ are hermitian
while $\Gamma_0$ is antihermitian. They satisfy
$$
\{ \Gamma_M, \Gamma_N \} = 2 g_{MN}. 
$$

\item{$\triangleright$}
$\Gamma_{M_1 \dots  M_n}$ is the antisymmetrized product of gamma matrices
$$
\Gamma_{M_1 \dots  M_n} =\Gamma_{[ M_1 \dots  } 
\Gamma_{M_n]}
$$
where the square bracket implies a sum over $n!$ terms with a $1/n!$ 
prefactor. 

\item{$\triangleright$}Gamma matrix identities that are useful are

$$
\eqalign{
& [ \gamma_m, \gamma^r]=2 {\gamma_m}^r \cr
& [ \gamma_{mnp}, \gamma^{rs} ]=12 {\delta_{[m}}^{[r} {\gamma_{np]}}^{s]}\cr 
& \{ \gamma_{mnpq}, \gamma^{rst} \} = 2 {\gamma_{mnpq}}^{rst} 
-72{\delta_{[mn}}^{[rs} {\gamma_{pq]}}^{t]}
}
$$

\item{$\triangleright$} Our definition of Hodge $\star$ in $d$ dimensions is

$$
\star ( dx^{m_1} \wedge \dots \wedge dx^{m_p} ) =
{| g|^{1/2}\over (d-p)!} 
{{\epsilon^{m_1 \dots m_p}}_{m_{p+1} \dots m_d} 
dx^{m_{p+1} }  \wedge \dots \wedge dx^{m_d} , 
}
$$
where 
$$
\epsilon_{m_1 m_2 \dots m_d}= 
\cases{0 & any two indices repeated \cr
+1 &  even permutation \cr 
-1 & odd permutation \cr }
$$

\item{$\triangleright$} 
The identity which relates covariant derivatives of spinors with 
respect to conformally transformed metrics is
$$
\eqalign{ 
& {\tilde \nabla}_M \epsilon = \nabla_M \epsilon 
+ {1 \over 2} \Omega^{-1} {\Gamma_M}^N ( \nabla_N \Omega) \epsilon , \cr
& {\tilde g}_{MN} =\Omega^2 g_{MN} . \cr}
$$

\listrefs

\end